\documentstyle[12pt,psfig]{article}

\makeatletter
\@addtoreset{equation}{section}
\makeatother

\renewcommand{\baselinestretch}{1.5}
\newcommand{\be}{\begin{equation}}
\newcommand{\ee}{\end{equation}}

\newcommand{\DS}{\renewcommand{\baselinestretch}{1.5} \tiny
  \normalsize }
\newcommand{\SS}{\renewcommand{\baselinestretch}{1} \tiny \normalsize}

\newcommand{\tta}{\widehat \tau}
\newcommand{\nin}{\in\!\!\!\!\!|}

\begin{document}
\centerline {\bf Nonparametric Methods for Doubly Truncated Data}
\centerline {Bradley Efron and Vah\'e Petrosian}

\begin{abstract}

Truncated data plays an important role in the statistical analysis of
astronomical observations as well as in survival analysis. The motivating
example for this paper concerns a set of measurements on quasars in
which there is double truncation. That is, the quasars are only
observed if their luminosity occurs within a certain finite interval,
bounded at both ends, with the interval varying for different
observations. Nonparametric methods for the testing and estimation of
doubly truncated data are developed. These methods extend some known
techniques for data that is only truncated on one side, in particular
Lynden-Bell's estimator and the truncated version of Kendall's tau
statistic. However the kind of hazard function arguments that underlie
the one-sided methods fail for two-sided truncation. Bootstrap and
Markov Chain Monte Carlo techniques are used here in their
place. Finally, we apply these techniques to the quasar data,
answering a question about their long-term luminosity evolution.
\end{abstract}

\vspace{2in}

\SS


\DS

\noindent {\bf Acknowledgement} \quad We are grateful to Susan Holmes,
Persi Diaconis, and Duncan Murdach for guidance on the MCMC methodology.

\DS

\section{Introduction}{\indent}

Figure 1 shows an example of doubly truncated astronomical data. The plotted points are
redshifts $z_i$ and log luminosities $y_i$   for n = 210 quasars, 
as explained in Section 6. Due to experimental constraints the
distribution of each $y_i$ is truncated to a known interval $R_i$
depending on $z_i$. Truncation means that we would not have learned of
$y_i$'s existence if it fell outside of region $R_i$. Many
experimental situations lead to truncated data, see for example
McLauren et al. (1991). The double truncation seen in Figure 1, where
$y_i$ goes undetected if it is either too small or too large, is
less common than one-sided truncation. The quasar data has still
another kind of truncation, of the redshift values $z_i$, but that
will not affect our discussion.

We will describe nonparametric methods that answer to two related questions
concerning truncated data:

\noindent {\bf Question 1} \quad How can we test whether or
not the $y_i$'s are independent of the $z_i$'s? 

\noindent {\bf Question 2} \quad Assuming independence, how can we
estimate the marginal distribution of the $y_i$'s?

\noindent The answers apply to all forms of truncation, including the
double truncation of Figure 1.  In fact some of our methods apply to
all forms of data censoring and truncation, as mentioned at the end
of Section 5. 

\begin{figure}[p]
\leavevmode\centering
\psfig{file=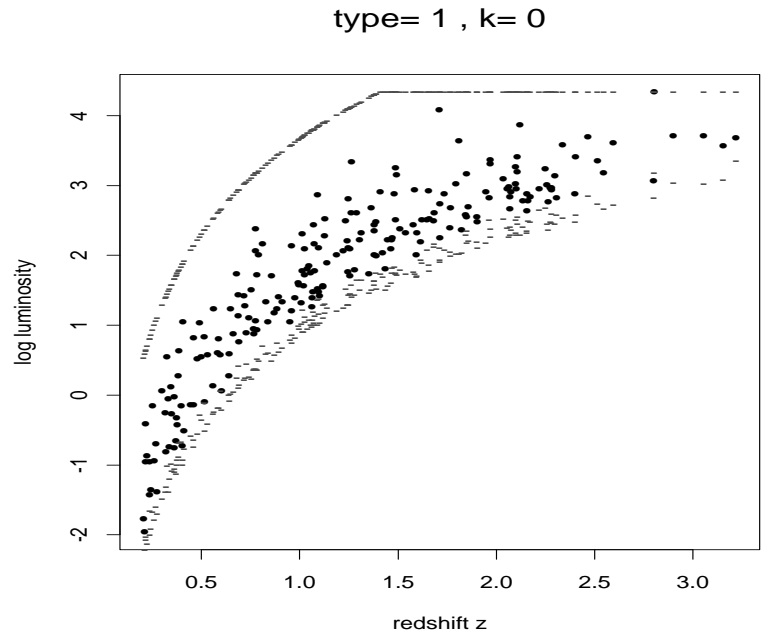,width=7.0in,height=7.0in}
\caption{Doubly truncated data; points
represent redshifts and luminosities for 210 quasars, as described in
Section 6; luminosity subject to lower and upper truncation as
indicated by dashes (lower truncation limits shifted down .25 for clarity.) Is
luminosity correlated with redshift?} 
\label{data}
\end{figure}

Question 1 turns out to be crucial for the scientific question
posed by the quasars. A positive dependence of luminosity on redshift
would mean that earlier quasars were intrinsically brighter, or in
other words that quasars have been evolving toward a dimmer state as
the universe ages.  Figure 1 certainly seems to show a strong positive
dependence between redshift and luminosity, but that appearance is
forced on us by the truncation limits, which increase sharply from
left to right. The tests described in Sections 2, 3, 5, and 6 will
show that a statistically significant positive relationship remains 
after accounting for truncation, though of much smaller magnitude than
the figure suggests.

Turnbull's important 1976 paper answers Question 2 for arbitrary
patterns of data truncation and censorship. Turnbull uses a
self-consistency algorithm to calculate the nonparametric maximum
likelihood estimate (MLE) of the $y$ distribution. His equations take
on interesting forms for doubly truncated data, as discussed in
Section 4.

Both Question 1 and Question 2 have simple closed-form solutions in
the {\it one-sided case}, where the truncation regions $R_i$
extend upwards to infinity. Tsai (1990) and also Efron and Petrosian
(1992, 1994) answer Question 1 using a version of Kendall's tau
statistic appropriate to one-sided truncation. Lynden-Bell's (1971)
answer to Question 2 in the one-sided case is closely related to the
Kaplan-Meier estimate for censored data, see Efron and Petrosian
(1994).

Sections 2-5 extend these ideas to two-sided truncation (the
extensions applying with little change to any pattern of multiple
truncation.)  The two-sided case is less tractable for both
questions.  The testing problem of Question 1 is particularly
challenging computationally, and raises issues concerning the
relationship of bootstrap methods to Markov Chain Monte Carlo (MCMC). 

Lynden-Bell's method is completely nonparametric, as is the tau test,
but it can be quite inefficient in some circumstances. Section 4 also
discusses more efficient parametric estimates based on {\it special
  exponential families}, as in Efron (1996) and Efron and Tibshirani (1996).

We return to the quasar data in Section 6. The testing and estimation
methods developed in Sections 2-5 are used there to answer Questions 1
and 2. 

\section{Permutation Tests for Independence}{\indent}

Doubly truncated data of the kind shown in figure 1 can be described as
follows: the observed data consists of $n$ pairs $(z_i, y_i)$, $y_i$ a
real-valued response and $z_i$ in covariate, with observation $y_i$
restricted to a known region $R_i = [u_i, v_i]$,

\be
{\rm data} = \{ (z_i,y_i) \quad {\rm with} \quad y_i \in R_i = [u_i, v_i]
\quad {\rm for} \quad i = 1, 2,\ldots, n \}. 
\ee

\noindent The $n$ quadruplets $(z_i, y_i, u_i, v_i)$ are observed
independently of one another. The regions $R_i$ can depend on $z_i$,
as in figure 1, and the $z_i$ values themselves can be subject to
observational truncation.

Question 1 concerns testing the {\it independence hypothesis} $H_o$:
that if we could observe the $y_i$'s without truncation they would be
independent and identically distributed (i.i.d.) according to some
common density function $f(y)$. Because of truncation, $H_o$ says that
$y_i$ has the conditional density of $f(y)$ restricted to $R_i$, 

\begin{eqnarray}
H_o: \quad f(y_i | R_i) &=& f(y_i) \slash F_i \quad {\rm for} \quad
y_i \in R_i \nonumber \\&=& 0 \quad {\rm for } \quad y_i  \nin \ \ R_i 
\end{eqnarray}

\noindent independently for $i = 1, 2,\ldots, n.$ Here

\be
F_i = \mathop\int_{u_i}^{v_i} f(y)dy.
\ee

\noindent Question 2 asks us to estimate $f(y)$ assuming that $H_o$ is
true.

This section discusses permutation tests of the independence
hypothesis $H_o.$ Table 1 and Figure 2 show an artificial example of
truncated data involving $n = 7$ points that will be helpful in
carrying out the discussion.

\renewcommand{\arraystretch}{.75}

\begin{table}[htbp]
  \begin{center}
    \leavevmode
    \begin{tabular}{cccccc}
      $z_i$ & $y_i$  & $R_i=[u_i,v_i]$ \\
      \hline
      1 & 0.75 & [0.4, 2.0] \\
      2 & 1.25 & [0.8, 1.8] \\
      3 & 1.50 & [0.0, 2.3] \\
      4 & 1.05 & [0.3, 1.4] \\
      5 & 2.40 & [1.1, 3.0] \\
      6 & 2.50 & [2.3, 3.4] \\
      7 & 2.25 & [1.3, 2.6]\\
    \end{tabular}
    \caption{Table 1:\quad An artificial example of truncated data involving $n = 7$
      data points.}
    \label{tab:regions}
  \end{center}
\end{table}

A permutation of ${\bf y} = (y_1, y_2, \ldots, y_n),$ say ${\bf
  y}^\ast = (y_1^*, y_2^*, \ldots, y_n^*)$
is {\it observable} if the permuted values all fall within their truncation
  regions, that is if

\be
y_i^\ast \in R_i \quad {\rm for} \quad i = 1, 2, \ldots, n
\ee

\noindent In figure 1, and Table 2, we see that $(y_3, y_2, y_1, y_4, y_5, y_6, y_7)$
is observable, but not $(y_2, y_1, y_3, y_4, y_5, y_6, y_7)$. We
define 

\be
{\cal Y} = {\rm set \; of \; observable \;  permutations.}
\ee

\noindent It turns out that ${\cal Y}$ has 78 members in the
seven-point example.

Permutation tests of the independence hypothesis $H_o$ are based on the
conditional distribution of ${\bf y}^\ast$ in ${\cal Y}$ given the
ordered values $y_{(1)}, y_{(2)}, \ldots, y_{(n)}$ of the observed
response vector ${\bf y}$, (assuming for convenience no ties), say 

\be
{\bf y}_{(\;)} = (y_{(1)}, y_{(2)}, \ldots, y_{(n)})
\ee

\noindent and also given the $z_i$ and $R_i$ values, 

\be
{\bf z} = (z_1, z_2, \ldots, z_n) \quad {\rm and} \quad {\bf R} =
(R_1, R_2, \ldots, R_n).
\ee

\begin{figure}[p]
\leavevmode\centering
\psfig{file=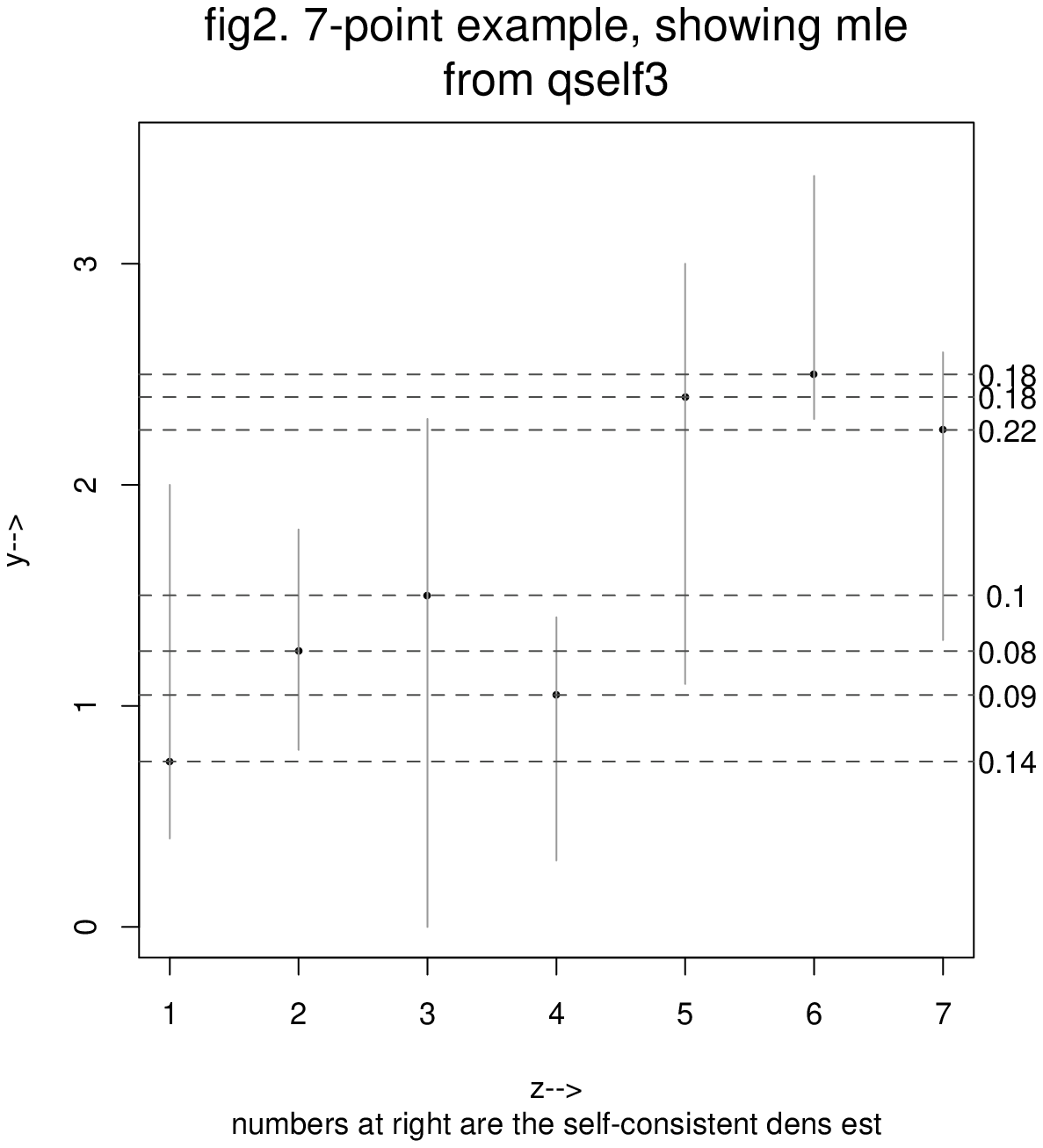,width=7.0in,height=7.0in}
\caption{The seven-point example from Table 1; dots
indicate points $(z_i, y_i);$ vertical bars show truncation regions
$R_i = [u_i, v_i];$ dashed horizontal lines indicate the ordered $y$
values. Numbers at right give the nonparametric MLE for the $y$
distribution assuming that the independence hypothesis $H_o$ is true,
as explained in Section 4.} 
\label{example}
\end{figure}

\noindent {\bf Independence Lemma}\quad Suppose that $H_o$ is
true. Then the conditional distribution of ${\bf y}^\ast$ given ${\bf
  y}_{(\;)}, {\bf z}, \ {\rm and} \ {\bf R}$ is uniform over
  ${\cal Y}$. 

\noindent {\it Proof}\quad According to (2.2), (2.3), an observable
  vector ${\bf y}^\ast = (y_1^\ast, y_2^\ast, \ldots, y_n^\ast)$ has
  $H_o$ density

\be
\prod_{i=1}^n [f(y_i^\ast)/F_i] = \prod_{i=1}^n f(y_i^\ast)/\prod_{i=1}^n F_i.
\ee

\noindent If ${\bf y}^\ast \in {\cal Y}$ then (2.8) equals
$\prod_{i=1}^n f(y_{(i)}) / \prod_{i=1}^n
  F_i.$ This has the same value for all ${\bf y}^\ast \in {\cal Y}$,
  which proves the lemma.

Permutation tests of $H_o$ are carried out in the usual way:  we
choose a test statistic $S({\bf y})$, with larger values of $S$
indicating stronger disagreement with $H_o$, and compare the observed
value $s = S({\bf y})$ with the set of permutation values

\be
{\cal S} = \{ S({\bf y}^\ast), {\bf y}^\ast \in {\cal Y} \}
\ee

\noindent The p-value of the test is the proportion of ${\cal S}$
exceeding $s$.

{\bf The tau test}.\quad A particular choice of the test statistic
$S ({\bf y})$ was advocated in Tsai (1990) and Efron and Petrosian
(1992, 1994), in the context of one-sided truncation. A pair of
indices $(i, j)$ is said to be {\it comparable} if $y_i \in R_j$ and
$y_j \in R_i$. Define 

\be
{\cal C} = {\rm set\;of\;comparable\;pairs}
\ee

\noindent and

\be
\widetilde \tau = \mathop \sum_{(i, j) \in {\cal C}} \; {\rm sign}[(y_i -
y_j)(z_i - z_j)] \ \slash \ \#{\cal C}.
\ee

\noindent For untruncated data $\#{\cal C} = \pmatrix{n \cr 2 \cr}$ and
$\widetilde \tau$ is Kendall's tau statistic. If we decide that the
independence hypothesis $H_o$ is false then $\widetilde \tau$ provides
a convenient nonparametric measure of correlation between $y$ and
$z$, see Section 3 of Efron and Petrosian (1994); $\widetilde \tau =
.429$ for the seven-point example. 

Here we will use just the numerator of (2.11),

\be
\widehat \tau = \sum_{(i, j) \in {\cal C}} {\rm sign} [(y_i - y_j)
(z_i - z_j)]
\ee

\noindent as the statistic $S({\bf y})$ for testing $H_o$, calling
this the {\it tau test}. For the seven-point example $\widehat \tau =
3$ and the 78 members of ${\cal Y}$ have this distribution of
$\widehat \tau^*$ values, 

\renewcommand{\arraystretch}{.75}
\begin{table}[htbp]
  \begin{center}
    \leavevmode
    \begin{tabular}{cccccc}
      $\widehat \tau^*$: & $< 3$ & $= 3$ & $> 3$ \\
      \#: & 63 & 8 & 7 \\  
     \end{tabular} \flushright (2.13)
    \label{tab:regions}
  \end{center}
\end{table}

\noindent The one-sided p-value for testing $H_o$ is $(7 + 8 / 2) /
78$ (splitting the probability atom at $\widehat \tau^* =
\widehat \tau$.) 

Permutation tests based on $\widetilde \tau$ or $\widehat \tau$ are
identical for one-sided truncation, since $\#{\cal C}$ is the same for
all observable permutations, but they can differ under two-sided
truncation. The two test statistics give almost the same results for
the quasar data. Section 6 also discusses other choices of $S({\bf
  y})$ that have greater testing power.

\section{Approximate P-Values}{\indent}

The p-value for the seven-point example was found by complete enumeration of the tau
statistic over the set of observable permutations ${\cal Y}$. This
becomes impossible for data sets much larger than $n = 7$, and we need
convenient approximations in order to carry out the test.

It seems like a good approximation should be easy to find. The tau
statistic (2.12) has permutation expectation zero,

\be
E_{perm} \{ \widehat \tau^\ast \} = 0,
\ee

\noindent since interchanging any comparable pair of indices reverses
the sign of the corresponding component of $\tta$ ((3.1) is not true
for $\widetilde \tau$, (2.11)). Moreover the components of $\tta$ are short-tailed so that the
normal approximation

\be
\tta \sim N(0, \sigma_{\rm perm}^2)
\ee

\noindent is likely to have good accuracy, see Theorem 2 of Tsai (1990). However,
there seems to be no convenient formula for $\sigma_{\rm perm}^2$, the
permutation variance, at least not in the doubly truncated case. A
formula does exist when the truncation is only one-sided, as discussed
below.

We can use {\it Markov Chain Monte Carlo} (MCMC) methods to
approximate $\sigma_{\rm perm}^2$. That is we can generate a random
walk on ${\cal Y}$ that eventually produces uniformly distributed
permutation vectors ${\bf y}^\ast$, and then estimate $\sigma_{\rm
  perm}^2$ by the empirical variance of $\tta({\bf y}^\ast)$. A
particularly simple MCMC scheme starts at ${\bf y}^\ast = {\bf y}$ and
proceeds as follows: \quad (1) choose a pair of indices ${i, j}$ at
  random; and \quad (2) interchange the $i^{th}$ and $j^{th}$
  components of ${\bf y}^\ast$ if the resulting vector is in ${\cal
    Y}$ (that is if $y_i^\ast \in R_j \quad {\rm and} \quad y_j^\ast \in
  R_i)$ otherwise keeping ${\bf y}^\ast$ the same. 

Elementary Markov Chain Theory says that ${\bf y}^\ast$ has its
stationary distribution uniform on ${\cal Y}$, see Gelman and Rubin
(1992). This assumes that ${\cal Y}$ is connected by coordinate
interchanges, a fact proved recently by Persi Diaconis and Ronald
Graham (personal communication). A sufficient number of iterations of steps (1) and (2) make
${\bf y}^\ast$ nearly uniformly distributed on ${\cal Y}$. Repeating
this whole process some large number $B$ of times, perhaps
$B = 800$, gives independent vectors ${\bf y}^\ast (1), {\bf y}^\ast
(2),\ldots, {\bf y}^\ast (B)$, distributed nearly uniformly over
${\cal Y}$. Then we can estimate the permutation variance by 

\be
\widehat \sigma_{\rm perm}^2 = \frac{\mathop \sum_{b=1}^B [\widehat
  \tau({\bf y}^\ast(b)) - \widehat \tau(\cdot)]^2}{B-1} \qquad \qquad \qquad
[\widehat \tau (\cdot) = \mathop \sum_{b=1}^B \widehat \tau ({\bf
  y}^\ast (b))/B],
\ee

\noindent and approximate the p-value of the tau test by 

\be
\widehat p = 1 - \Phi (\widehat T), \qquad \qquad \qquad \widehat T =
\tta ({\bf y})/\widehat \sigma_{\rm perm}
\ee

\noindent where $\Phi$ is the standard normal cumulative distribution
function (CDF). A more direct approach, but one that tends to require
larger values of $B_1$ is to estimate $p$ as at (2.9), by 

\be
\widehat p = \# \{ \widehat \tau^* > \widehat \tau \} / B \; .
\ee

An MCMC analysis of the quasar data of Figure 1 was carried out by
Duncan Murdoch, personal communication. The number of iterations
required to compute $\widehat T(0) = \widehat \tau / \widehat
\sigma_{\rm perm}$ with the same acuracy achieved in Figure 5 was,
{\it very} roughly, 4,000,000. In any given situation it is hard to
know how many iterations of steps (1) and (2) are required to make
${\bf y}^*$ sufficiently uniform on ${\cal Y}$. Generating independent
replicates ${\bf y}^*(b)$ as in (3.3), which is very convenient
for error analyses, may be quite inefficient in the MCMC context. An
information reference on these difficulties and their possible
remedies is Gilks et al. (1993) and the ensuing discussion.

Section 5 discusses a bootstrap approximation for $\widehat
\sigma_{\rm perm}^2$ that sacrifices the theoretical exactness of the
MCMC algorithm for more efficient and more definite numerical
results. Bootstrap estimates are used in section 6's analysis of the
quasar data. The bootstrap approach has the advantage of applying to
any kind of truncated and/or censored data, as mentioned at the end of
Section 5.

Neither MCMC nor the bootstrap are needed in the case of {\it
  one-sided truncation} where the observational regions in (2.11) are
  all of the form $R_i = [u_i, \infty)$. In this case there is a
  simple way to generate vectors ${\bf y}^\ast$ uniform on ${\cal Y}$.
Define the {\it risk-set numbers}

\be
N_j = \# \{ i: \; u_i \leq y_{(j)} \qquad {\rm and} \qquad y_i \geq
y_{(j)} \}
\ee

\noindent $N_j$ is the size of the $j^{th}$ risk set in the following sense:
Looking at figure 2, but with all the regions $R_i$ now extending up
to infinity, we begin with $y_{(1)}$,  the smallest
$y$ value, and work upwards. At each $y_{(j)}$ there are $N_j$
observable choices of ``$z_i$'' to go with $y_{(j)}$, all of which are
equally likely under $H_o$. These are the values of $z$ ``at risk'' for
pairing with $y_{(j)}$. For example, there are $N_2 = 3$ possible
choices of $z$ to go with $y_{(2)}$, namely 2, 3, or 4, the actual
choice in figure 2 being $z = 4$. Altogether there are $N =
\prod_{i=1}^n N_i$ members of ${\cal Y}$, $N = 324 \quad (=
3\cdot3\cdot3\cdot3\cdot2\cdot2\cdot1)$ for the seven-point example.

A uniform choice of ${\bf y}^\ast$ in ${\cal Y}$ is accomplished by
choosing $z_i$ uniformly from the $N_i$ possible choices at each
$y_{(j)}$. this makes it easy to simulate the permutation distribution
for any test statistic $S({\bf y})$. The permutation variance of the
tau statistic (2.12) turns out to be

\be
\sigma_{\rm perm}^2 = 4 \sum_{i=1}^n V_i 
\qquad {\rm where} \qquad V_i = \frac{N_i^2 - 1}{12}
\ee

\noindent as shown for example in Section 3 of Efron and Petrosian
(1994). We will use Formula (3.7) in Section 5 to validate the accuracy
of the bootstrap estimate of $\sigma_{\rm perm}^2$. 

\section{Estimating The Response Distribution}{\indent}

Question 2 of the introduction asks us to estimate the distribution of the
response variable $y$ assuming that the independence hypothesis $H_o$
is true. More precisely, we want to estimate the response density
$f(y)$ in (2.2), (2.3). This section discusses nonparametric and
parametric estimates of $f(y)$ when the data is doubly truncated.

The nonparametric MLE is a discrete distribution putting all of its
probability on the observed responses $y_1, y_2, \ldots, y_n$,
Turnbull (1976). Let ${\bf f} = (f_1, f_2, \ldots, f_n)$ be a
distribution putting probability $f_i$ on $y_i$, and let ${\bf F}
= (F_1, F_2, \ldots, F_n)$ be the vector of observational
probabilities $F_i = {\rm prob}_{\bf f} \{ y \in R_i \}$, so 

\be
{\bf F} = {\bf J}{\bf f}
\ee

\noindent where ${\bf J}$ is the $n \times n$ matrix describing which
$y$ values are included in the regions $R_i$, 

\begin{equation}
J_{ij} = \left\{ \begin{array}{cc}
    1 & \mbox{if $y_j \in R_i$}\\
    0 & \mbox{if $y_j \nin \;\;R_i$}
    \end{array}
\right.
\end{equation}

According to (2.2), the log likelihood of the observed sample is

\be
\ell = \log \prod_{i=1}^n (f_i/F_i)
\ee

\noindent Differentiating (4.3) with respect to $f_i$, and using
(4.1), gives 

\be
\frac{\partial \ell}{\partial f_i} = \frac{1}{f_i} - \sum_{j=1}^n
J_{ji} \frac{1}{F_j}.
\ee

\noindent The maximum likelihood equations $\partial \ell / \partial
f_i = 0$ can be expressed as 

\be
\frac{{\bf 1}}{{\bf f}} = {\bf J}^\prime \frac{{\bf 1}}{{\bf F}},
\ee

\noindent where $\frac{{\bf 1}}{{\bf f}} = (\frac{1}{f_1},\frac{1}{f_2},
\ldots, \frac{1}{f_n})\quad {\rm and} \quad \frac{{\bf 1}}{{\bf F}} =
(\frac{1}{F_1}, \frac{1}{F_2}, \ldots, \frac{1}{F_n})$. Notice that
$\ell$ in (4.3) stays the same if ${\bf f}$, and hence ${\bf F}$, is
multiplied by any positive constant, allowing us to ignore the
constraint $\sum_{i=1}^n f_i = 1$ in the derivation of (4.5).

Equation (4.5) is the same as Turnbull's self-consistency
criterion. We can solve for the MLE $\widehat{\bf f}$ by beginning
with any initial estimate and then iterating between (4.1) and (4.5)
(remembering to re-scale after each application of (4.5) so that the
estimate of ${\bf f}$ sums to 1). The nonparametric MLE $\widehat {\bf
  f}$ for the seven-point example is shown at the right edge of Figure
2. The substantial differences between $\widehat {\bf f}$ and the
untruncated MLE (.14, .14, ..., .14) are not intuitively obvious from
the truncation pattern.

The method just described is an EM algorithm, and often converges
quite slowly to the MLE. An alternative algorithm is based on
Lynden-Bell's 1971 method for computing the MLE in the case of
one-sided truncation. For notational convenience suppose that the $y$
values are indexed in increasing order, so $y_i = y_{(i)}$, where we
continue to assume that there are no ties. Corresponding to density
function ${\bf f} = (f_1, f_2, \ldots, f_n)$, the {\it survival curve} ${\bf G}
= (G_1, G_2, \ldots, G_n)$ and the {\it hazard function} ${\bf h} =
(h_1, h_2, \ldots, h_n)$ are defined by 

\be
G_j = \sum_{i \geq j} f_i \qquad {\rm and} \qquad h_j = f_j/G_j
\ee

\noindent We can recover ${\bf G}$ and ${\bf f}$ from ${\bf h}$ via
the relationships 

\be
G_j = \exp \{ \sum_{i<j} \log (1-h_i) \} \qquad {\rm and} \qquad f_j =
G_j - G_{j+1}
\ee

\noindent Here $G_1 = G(y_{(1)}) = 1$ by definition.

The following theorem is verified in the Appendix.

\noindent {\bf Hazard Rate Theorem}\quad The nonparametric MLE
$\widehat {\bf f}$ has hazard function $\widehat {\bf h}$ satisfying

\be
\frac{1}{\widehat h_j} = N_j + \sum_{i=1}^n J_{ij} \widehat Q_i
\ee

\noindent where $N_j$ are the risk-set numbers (3.5), $\{ J_{ij} \}$ are
the inclusion indicators (4.2), and 

\be
\widehat Q_i = \widehat G_{v_i+} \ \slash \ \widehat F_i
\ee

\noindent Here $\widehat{\bf F} = {\bf J}\widehat{\bf f}$ as in (4.1),
and $\widehat G_{v_i +} = \sum \{ \widehat f_k : y_k > v_i \}$.

The numerator of $\widehat Q_i$ is the MLE probability of
exceeding $v_i$, the upper observational limit for $y_i$. In the
one-sided truncation case $\widehat Q_i = 0 \quad {\rm since}
\quad v_i = \infty$, so equation (4.8) takes the form

\be
\frac{1}{\widehat h_j} = N_j \ ,
\ee

\noindent which is Lynden-Bell's (1971) estimate. In this case (4.7) gives the
MLE $\widehat{\bf f}$ directly, without iteration. 
In the case of two-sided truncation we can begin with
(4.10) and iterate (4.7), (4.8) to obtain the MLE. This converges
quickly if the upper truncation is not severe, as turns out to be the
situation for the quasar data.

The solid curve in Figure 3 shows $\log \widehat G(y)$, the
nonparametric MLE of $\log (\sum_{y_i\geq y}f_i)$, for the 210
quasars. The log luminosity $y$ is not the same quantity plotted in
figure 1, but is an adjusted version as explained in Section 6. Also
\noindent shown is the Lynden-Bell estimate (4.10), dashed line, 
which ignores upper truncation. The two estimates are almost the same so in this case
upper truncation has little effect. The MLE algorithm based on
(4.7)-(4.8) converges very quickly here.

\begin{figure}[p]
\leavevmode\centering
\psfig{file=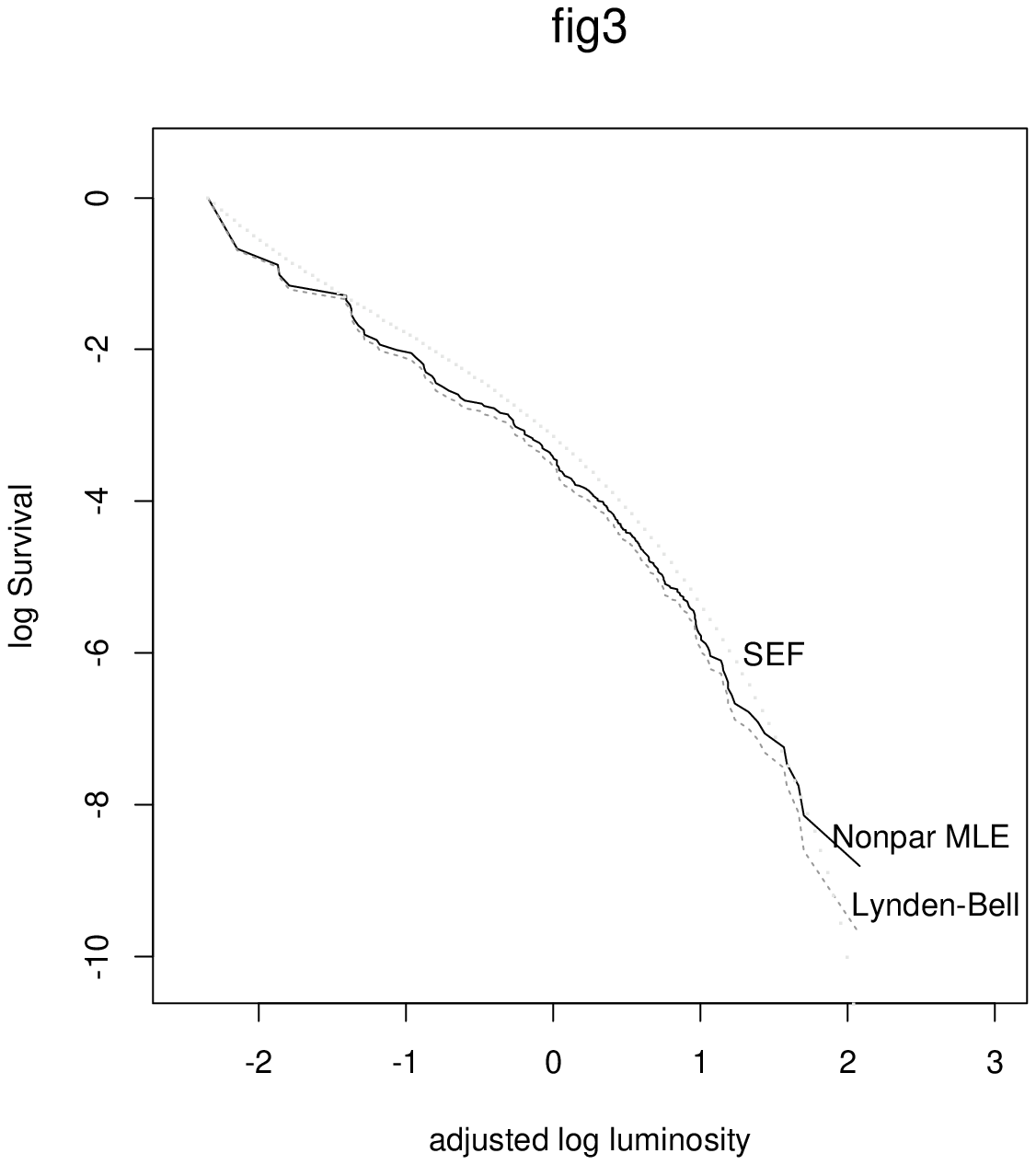,width=7.0in,height=7.0in,angle=0}
\caption{Estimated log survival curve $\log G(y)$ as a
function of the adjusted log luminosity evolution $y_j$ for the quasar data,
adjusted for luminosity evolution:  $\theta = 2$, as
explained in Section 6; solid curve is nonparametric MLE; dashed curve
is Lynden-Bell estimate (4.10), ignoring upper truncation; smooth
dotted curve is cubic special exponential family.} 
\label{LF}
\end{figure}

It is not an accident that the Lynden-Bell estimate of survival 
is everywhere less than the MLE.

{\bf Corollary}\quad The Lynden-Bell estimated survival curve, which
ignores upper truncation, is less than or equal to the nonparametric
MLE. 

The proof of the corollary is immediate from a comparison of
(4.8) with (4.10), which shows that the estimated hazard rate can only
be greater in the Lynden-Bell case.

The curve labeled ``SEF'' in Figure 3 is derived from a {\it special
  exponential family} in the terminology of Efron and Tibshirani
  (1996). In this case the SEF density estimate $\widehat f(y)$ is the
  MLE among densities of the form

\be
\log f(y) = \beta_0 + \beta_1y + \beta_2y^2 + \beta_3y^3,
\ee

\noindent for $y$ in the range of the observed $y_i$ values. That is,
it maximizes $\prod_{i=1}^n (f(y_i)/F_i)$ in (2.2), (2.3) among all
choices of $(\beta_1, \beta_2, \beta_3)$ (with $\beta_0$ then
determined by the requirement that $f(y)$ integrate to 1 over the
interval $[y_{(1)}, y_{(n)}]$). The appendix describes the calculation
of $\widehat f(y)$. 

Lynden-Bell's estimate and the parametric MLE can behave erratically
near the extremes of the $y$ range, where the risk set numbers $N_j$
may be small. At the arrowed point in figure 3 for example $N_1 =
2$, giving Lynden-Bell estimates $\widehat h_1 = .50 \quad {\rm and}
\quad \widehat G(y_{(2)}) = .50$ according to (4.10) and (4.7). The
nonparametric MLE has $\widehat G(y_{(2)}) = .51$. The SEF estimate
smooths out the bumps in the nonparametric MLE, here giving $\widehat
G(y_{(2)}) = .72$. 

SEF estimates are less variable than the nonparametric MLE, but can be
biased if based upon an incorrect model. The cubic model in Figure 3
was chosen by successive significance tests, as explained in the
appendix. A small simulation study showed that percentile points
of the cubic SEF estimates had roughly half the standard deviation of
their nonparametric MLE counterparts.

\section{Bootstrap Tests for Independence}{\indent}

We now return to the question of testing the independence hypothesis $H_o$, (2.2). this
section discusses bootstrap approximations to the permutation p-values
of Section 2. Most of the discussion is in terms of the tau test, but
the method applies to any test statistic as mentioned at section's
end, and can be extended to arbitrarily complicated patterns of data
censoring and truncation.

Let $\widehat f(y)$ be an estimate of the density for the response
variable $y$ calculated as in Section 4, assuming that $H_o$ is
true. The estimate $\widehat f$ might be the nonparametric MLE or an 
SEF estimate as in Figure 3. We can use $\widehat f$ to draw a
{\it bootstrap sample} ${\bf y}^\ast$ by following the recipe in
(2.2), (2.3):\quad ${\bf y}^\ast = (y_1^\ast,
y_2^\ast,\ldots,y_n^\ast)$ has independent components, with the
$i^{th}$ component's density being 

\be 
\left.
\begin{array}{cl}
\widehat f(y_i^\ast)/\widehat F_i & {\rm for} \ y_i^\ast \ \in \ R_i \\ 
0 & {\rm for} \ y_i^\ast \ \nin \ R_i
\end{array} 
\right\} 
\ {\rm independently} \ i = 1, 2, \ldots, n 
\ee

\noindent where $\widehat F_i = \int_{R_i} \widehat f(y)dy$. 

An approximate version of the tau test can be carried out by
generating $B$ independent bootstrap samples ${\bf y}^\ast$ and then
proceeding as in (3.3), (3.4) to get the approximate p-value $\widehat
p = 1 - \Phi (\widehat T)$. How big need $B$ be? Letting 

\be
T = \widehat \tau ({\bf y})/ \sigma_{\rm perm} \quad {\rm and} \quad
\widehat T = \widehat \tau ({\bf y}) / \widehat \sigma_{\rm perm},
\ee

\noindent standard normal-theory calculations show that 

\be
sd_\ast \{ \widehat T / T \} \doteq 1/\sqrt {2B},
\ee

\noindent where $sd_\ast$ indicates the simulation standard
deviation. This gives 

\renewcommand{\arraystretch}{.75}
\begin{table}[htbp]
  \begin{center}
    \leavevmode
    \begin{tabular}{cccccc}
      B: & 50  & 100 & 200 & 400 & 800  \\
      $sd_\ast$ & 0.10 & .07 & .05 & .035 & .025  \\ 
      \end{tabular} \flushright (5.4)
    \label{tab:regions}
  \end{center}
\end{table}

\noindent so we need $B = 800$ bootstrap replications to determine
$\widehat T$ within about 2.5\% of its ideal value $T$. These same
calculations apply to the MCMC approach in (3.3), (3.4). $B = 800$ is
large enough to permit a check in the normality assumption leading to
the estimate $\widehat p = 1 - \Phi (\widehat T)$, and a retreat to
the nonparametric estimate (3.5) if normality looks dubious. 

As a check on the bootstrap test, $\widehat \sigma_{\rm perm}$ was
estimated using $B = 800$ replications for the data in Figure 1, but
ignoring the upper bounds of the truncation regions (i.e. taking all
$v_i = \infty$). The estimate was $\widehat \sigma_{\rm perm} = .0441 \pm
.0011$, ``$\pm$'' indicating the bootstrap simulation error, agreeing
nicely with the exact permutation standard deviation $\sigma_{\rm
  perm} = .0445$ obtained from (3.7).

The bootstrap approximation for $\sigma_{\rm perm}$ looks much
different than the MCMC approach of Section 3. In particular it
requires a preliminary estimate of $f(y)$. The brief discussion in the
appendix shows that the bootstrap and permutation calculations are
more alike than they appear, and that similar approximations are 
used in more familiar statistical contexts.

There is nothing special about the tau statistic as far as bootstrap
or permutation methods are concerned. Section 6 mentions a more
powerful version of tau that puts greater weight on those terms in
(2.11) having bigger values of $| z_i - z_j |$. We could in fact
employ any test statistic $S({\bf y})$, and
then use the bootstrap or MCMC to approximate the comparison set
${\cal S}$ and p-value in (2.9). The main advantage of tau-like test
statistics is that they have bootstrap or permutation expectations
zero under $H_o$, and nearly normal distributions, which eases the
task of finding p-values by simulation.

Tsai (1990, Section 4) shows that the tau test of independence can be
applied to data that is truncated below and censored above. In
principle, bootstrap tests can be applied to any form of truncated
and/or censored data. The first step is to estimate the response
density $f(y)$ using Turnbull's (1976) nonparametric self-consistency
algorithm. Then bootstrap samples ${\bf y}^\ast$ are drawn as in (5.1),
taking into account each $y_i$'s pattern of truncation and censoring,
and the p-value approximated by comparing the test statistic
$S({\bf y})$ with ${\cal S} = \{ S({\bf y}^\ast (1)), S({\bf y}^\ast
(2)), \ldots, S({\bf y}^\ast (B)) \}$. Romano (1988) gives a general
discussion, and validation, of this kind of ``null hypothesis
bootstrap'' test procedure.

\section{The Quasar Data}{\indent}

Our estimation and testing theory will now be applied to the quasar data of Figure 1. First the
situation will be described more carefully. The original dataset
consisted of independently collected quadruplets

\be
(z_i, m_i, a_i, b_i) \quad i = 1, 2, \ldots, n
\ee

\noindent where $z_i$ is the redshift of the $i^{th}$ quasar and $m_i$
is its apparent magnitude. The numbers $a_i$ and $b_i$ are lower and upper truncation limits on
$m_i$. Quasars with apparent magnitude above $b_i$ were too dim to yield
dependable redshifts (remembering that bigger values of $m$ correspond
to dimmer objects.) The lower limit $a_i$ was used to avoid confusion
with non-quasar steller objects. In this
study $a_i = 16.08$ for all $i$, while $b_i$ varied between 18.494 and
18.934. The full dataset comprised $n = 1052$ quasars. Here we are
considering a randomly selected subset of $n = 210$ quadruplets
(6.1). 

Farther quasars appear dimmer of course, that is they tend to have
bigger values of $m_i$. Hubble's law, which says that distance is
proportional to redshift, allows us to transform apparent magnitudes
into a luminosity measurement that should be independent of
distance. This transformation depends on the cosmological model
assumed. The log luminosity values $y_i$ in Figure 1 were obtained
from a formula $y_i = t(z_i, m_i)$ that takes into account relativistic
effects of the distance, 

\be
y_i = 19.894 - 2.303 \frac{m_i}{2.5} + 2 \log (Z_i -
  Z_i^{1/2}) - \frac{1}{2} \log (Z_i),
\ee

\noindent where $Z_i = 1 + z_i$. Formula (6.2) is derived from the
Einstein-deSitter cosmological model, Weinberg (1972). The last term
makes the so-called K-correction, taking into account the shifting of
the spectrum due to redshift.

Larger values of $y_i$ correspond to intrinsically brighter quasars in
Figure 1. The truncation limits $R_i = [u_i, v_i]$ were obtained by
applying transformation (6.2) to the observational limits for $m_i$,

\be
u_i = t(z_i, b_i) \quad {\rm and} \quad v_i = t(z_i, a_i)
\ee

\noindent This makes $u_i$ and $v_i$, the lower and upper dashes in
figure 1, strongly increasing functions of $z_i$, even though $a_i$ and $b_i$ are not.

 One of the principal goals of the quasar investigations is to study {\it
  luminosity evolution}: quasars may have been intrinsically brighter
  in the early universe and evolved toward a dimmer state as time went
  on. This would tend to make the points on the right side of Figure 1
  higher since larger redshifts correspond to quasars seen longer
  ago. However in the absence of luminosity evolution we should have $y_i$
  independent of $z_i$ except for truncation effects. This brings us
  back to Question 1 of the introduction. Testing the independence
  hypothesis $H_o$ amounts to testing for the absence of luminosity
  evolution.

A convenient one-parameter model for luminosity evolution says that
the expected log luminosity increases linearly as $\theta \cdot \log(1 +
z)$, with $\theta = 0$ corresponding to no evolution. If $\theta$ is
a hypothesized value of the evolution parameter then instead of $y_i$
being independent of $z_i$ we should test the null hypothesis
``$H_\theta$'', 

\be
H_\theta: \qquad y_i(\theta) = y_i - \theta \cdot \log(1+z_i) \qquad {\rm
  independent \; of} \quad  z_i
\ee

\noindent Figure 4 shows plots of the data for $\theta$ equal 0, 2,
and 4, all for our same set of 210 quasars. Notice that the truncation
regions $R_i = (u_i, v_i)$ also change with $\theta$, 

\be
u_i(\theta) = u_i - \theta \cdot \log (1+ z_i) \quad {\rm and} \quad
v_i(\theta) = v_i - \theta \cdot \log (1+z_i)
\ee

\begin{figure}[p]
\leavevmode\centering
\psfig{file=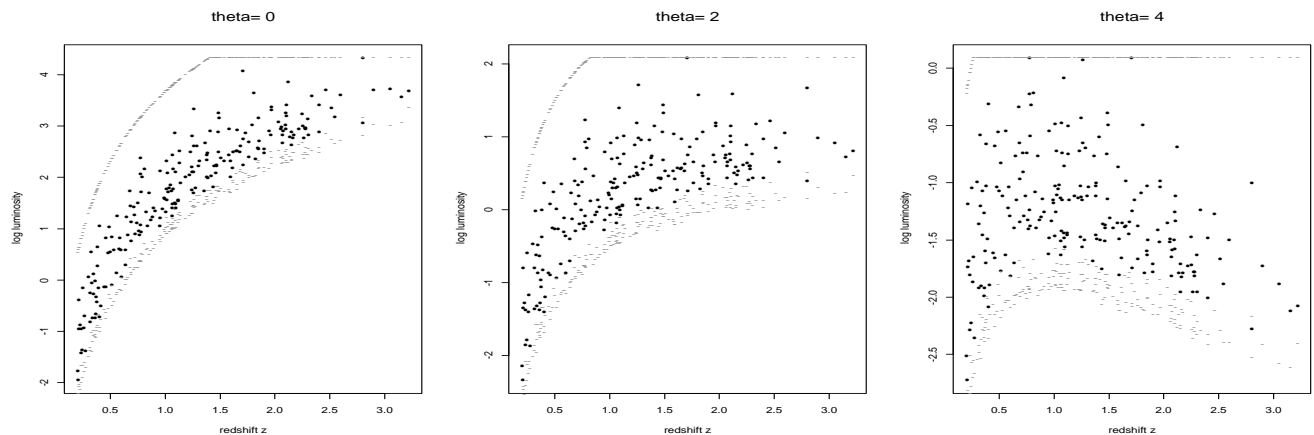,width=7.0in,height=7.0in,angle=0}
\caption{Plots of the quasar data of Figure 1 for
three choices of the luminosity evolution parameter $\theta$; $\theta
= 0$ corresponds to Figure 1; other values of $\theta$ plot $(z_i,
y_i(\theta)),$ (6.4), with limits $u_i(\theta)$ and $v_i(\theta)$,
(6.5).} 
\label{LumEvol}
\end{figure}

We can apply the tau test to each of the null hypotheses
$H_\theta$. This was done for values of $\theta$ between 0 and 4 with
the results shown in figure 5. The solid curve is the 
standardized test statistic $\widehat
T = \widehat \tau \slash \widehat \sigma_{\rm perm}$, (3.4), with
$\widehat \sigma_{\rm perm}$ determined by bootstrap sampling. $B =
800$ bootstrap replications (5.1) were drawn for $\theta = 0, .5, 1,
\ldots, 4$. Almost exactly the same curve was obtained using
$\widetilde \tau,$, (2.11), in place of $\widehat \tau$. 

\begin{figure}[p]
\leavevmode\centering
\psfig{file=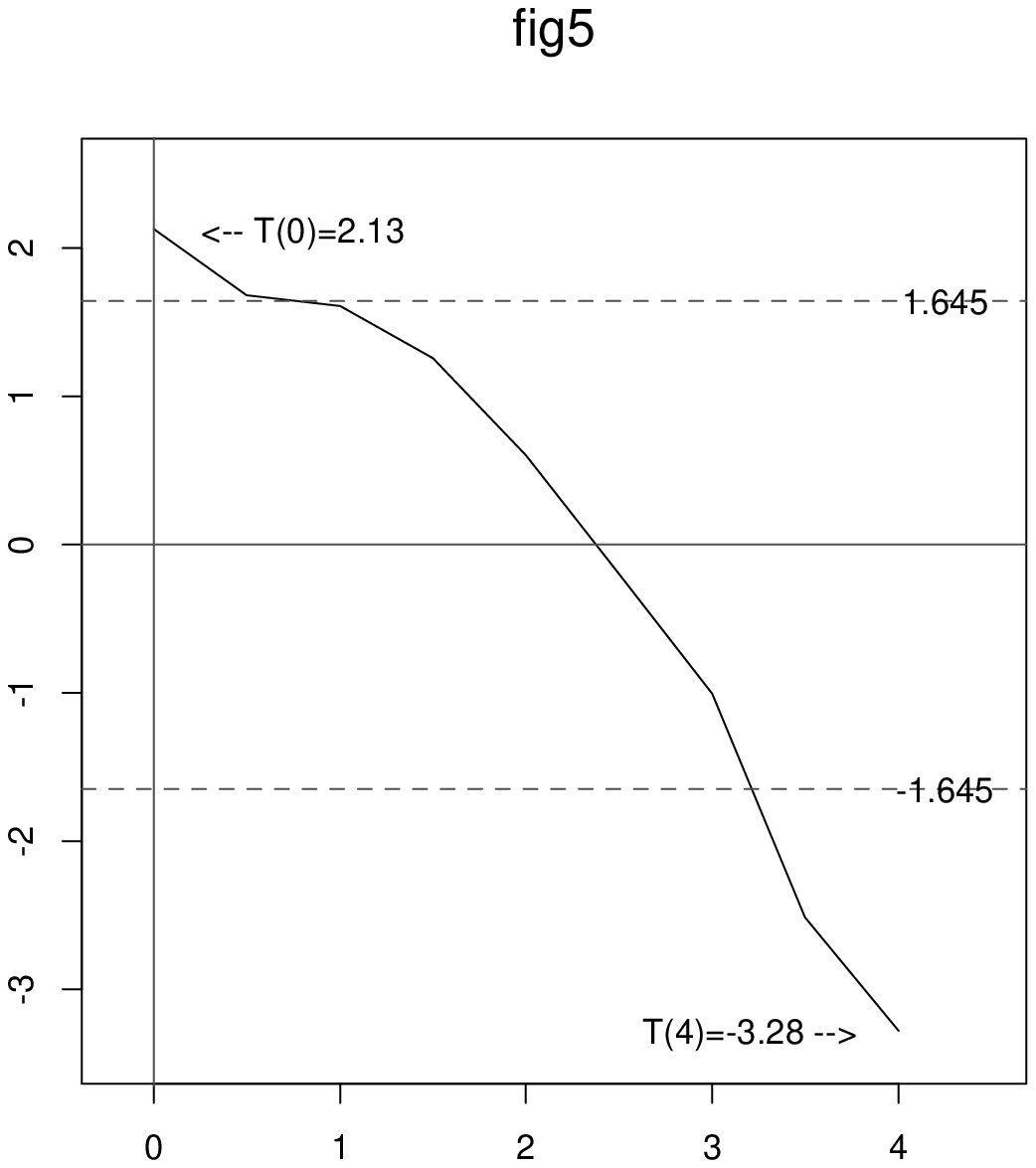,width=7.0in,height=7.0in}
\caption{Tau test of the null hypotheses
$H_\theta$, (6.4), for the 210 quasars; $\theta = 0, 0.5, 1, \ldots,
4$; solid curve $\widehat T = \widehat \tau \slash \widehat
\sigma_{\rm perm}$, (3.4), crosses zero at $\widehat \theta = 2.38; \;
\widehat \sigma_{\rm perm}$ found by bootstrap sampling (5.1); 90\%
central confidence interval $\theta \in [1.00, 3.20]$. Dots indicate
$\widehat T$ values assuming only one-sided truncation.} 
\label{tau}
\end{figure}

We see that $\widehat T(0) = 2.20$, giving an approximate one-sided
p-value $1 - \Phi \; (2.13) = .015$. The tau test rejects the null
hypothesis of independence $H_o$ in favor of a positive value of
the luminosity parameter $\theta$. At $\widehat \theta = 2.38$
we have $\widehat T(\widehat \theta) = 0$. The $\widehat T(\theta)$
curve crosses $\pm 1.645$ at [1.00, 3.20], which provides an
approximate 90\% central confidence interval for $\theta$. As a point
of comparison, using all 1052 quasars gave $\widehat \theta = 2.11$
and 90\% interval [1.38, 2.63].

If we are willing to ignore the upper truncation limits (setting all
the $v_i = \infty$) we can employ the more exact one-sided tau test
(3.5)-(3.6). These results, which did not involve bootstrap sampling,
were only slightly more significant than those for the two-sided test,
as shown by the dots in Figure 5.

The choice $\theta = 2$ makes the adjusted log luminosity $y_i(2) =
y_i - 2 \log (1 + z_i)$ approximately independent of $z_i$ so we can
estimate its density $f(y)$ as in Section 4. Figure 3 shows the
estimated log survival curve $\log \widehat G$. It curves sharply
downward for large values of $y$, which is a fortunate thing for the
testing results. If $\log \widehat G$ were linear then $f(y)$ would
correspond to a one-sided exponential density. Because of the
exponential's memoryless property {\it it is impossible
  to test for independence in the exponential case} (unless the lower
endpoint of the exponential density exceeds some of the lower
truncation limits $u_i$).

On the other hand, the MLE $\widehat f(y)$ consistently estimates an
exponential tail even if $y_i \quad {\rm and} \quad z_i$ are not
independent. It is a good idea to estimate $f(y)$ or $G(y)$ even if
Question 1 is of primary interest. Should the MLE turn out to be
exponential then any testing procedure will be futile. 

Model (6.4) gives us reason to question the efficacy of the tau
statistic in this situation. Let $\theta_o$ be the true value of the
evolution parameter. Then the difference of two of the luminosity
measurements in Figure 1, where $\theta = 0$, can be expressed as 

\be
y_i - y_j = y_i (\theta_o) - y_j(\theta_o) + \theta_o(w_i - w_j)
\qquad [w_i = \log (1+z_i)]
\ee

\noindent The differences $y_i(\theta_o) - y_j(\theta_o)$ are
symmetrically distributed about zero, except for truncation effects
so (6.6) suggests that $(i, j)$ pairs with bigger values of $|w_i -
w_j|$ will be more informative in testing for deviations from
$H_o$. An analysis of the 4907 comparable pairs for the data in Figure
1 verified that those with large values of $|w_i - w_j|$ were
contributing more consistently to the tau statistics:

\renewcommand{\arraystretch}{.75}
\begin{table}[htbp]
  \begin{center}
    \leavevmode
    \begin{tabular}{cccccc}
      $|w_i - w_j|: $ & 0 & .25 & .5 & .75 & 1 \\
      ${\rm prob} \{ {\rm sign} (y_i - y_j)(z_i - z_j) = 1 \}$ & .51 &
      .56 & .59 & .62 & .63 \\
     \end{tabular}
    \label{tab:regions}
  \end{center}
\end{table}

A modified version of the tau statistic (2.11) was tried on the quasar
data,

\be
\tilde \tau^\prime = \sum_{\cal C} | \log (w_i/w_j) \ | \ {\rm sign} \  [(y_i -
y_j)(z_i - z_j)] \ \slash \ \sum_{\cal C} | \log (w_i \slash w_j) |, 
\ee

\noindent but it showed only modest improvements over either $\widetilde
\tau$ or $\widehat \tau$. The gains were more substantial when all 1052 quasars were
included, giving standardized statistic $\tilde T^\prime = 3.21$ compared
to $\widehat T = 2.77$ for testing $H_o$.

\clearpage

\appendix
\centerline {\bf Appendix}

\section{Proof of the hazard rate theorem (4.8)}{\indent}\quad Because
the nonparametric MLE puts all of its probability on the observed
responses, and because it is invariant under monotonic transformation
of the $y$ scale, we can assume that $y_i = i$ for $i = 1, 2,\ldots$,
and that $R_i = [u_i, v_i]$ where $u_i$ and $v_i$ $\in \{ 1, 2,
\ldots, n \}$ (again assuming no tied $y$ values.) In addition to $N_j
= \# \{ i: \ u_i \leq j$ and $i \geq j \},$ (3.5), we define

\be
M_j = \# \{i: \ u_i = j \}
\ee

It is easy to see that 

\be
M_1 = N_1 \quad {\rm and} \quad M_j = N_j - N_{j-1} + 1 \quad {\rm
  for} \ \ j = 2, 3,\ldots, n.
\ee

\noindent We will also assume that 

\be
M_n = 0, 
\ee

\noindent because if this were not true then the largest observation
$y_i = n$ would have been truncated to the degenerate interval $R_n =
[n, n]$, and we could reduce the sample size to $n - 1$. (A3)
implies $N_{n - 1} = N_n + 1 = 2$ according to (A2), since $N_n = 1$. 

Following the notation of section 4, a density ${\bf f} = (f_1, f_2,
\ldots f_n)$ has likelihood $L = \prod_{i=1}^n f_i / F_i$ for
the observed sample. The hazard function ${\bf h} = (h_1, h_2, \ldots,
h_n)$ has $h_n = 1$, so we need only consider its first $n - 1$
coordinates. The following lemma expresses the likelihood in terms of
the hazard rate.

\noindent {\bf Lemma}\quad The likelihood corresponding to $(f_1,
f_2,\ldots, f_n)$ or $(h_1, h_2, \ldots, h_{n - 1})$ is 

\be
L = \prod_{i=1}^n \frac{f_i}{F_i} = [\prod_{j = 1}^{n - 1} h_j (1 -
h_j)^{N_j - 1}] \  / \ [\prod_{i = 1}^n \{ 1 - \prod_{u_i}^{v_i} (1 - h_k)
\}] 
\ee

\noindent {\it Proof}\quad In the one-sided case of no upper
truncation, where all the $v_i = n$, the denominator in (A4) equals
1 (since $h_n = 1$), so that (A4) becomes

\be
\prod_{i=1}^n \frac{f_i}{F_i} = \prod_{j=1}^{n - 1} h_j (1 - h_j)^{N_j
  - 1} 
\ee

\noindent This is easy to verify, starting from $h_j = f_j / G_j$:

\begin{eqnarray}
\prod_{j = 1}^{n - 1} h_j(1 - h_j)^{N_j - 1} &=& \prod_{j = 1}^{n - 1}
\frac{f_j}{G_j}(\frac{G_{j + 1}}{G_j})^{N_j - 1} = [\prod_{j = 1}^{n -
  1} f_j] \ [\frac{G_n^{N_{n - 1} - 1}}{G_1^{N_1},G_2^{N_2 + 1} \ldots
    G_{n - 1}^{N_{n - 1},-N_{n - 2} + 1}}] \\
&=& \prod_{j = 1}^n f_j / \prod_{j = 1}^{n - 1} G_j^{M_j},
\end{eqnarray}

\noindent where we have used $G_n = f_n, \quad N_{n - 1} = 2,$ (A2),
and (A3). But in the one-sided case $\prod_{j = 1}^{n - 1} G_j^{M_j} =
\prod_{i = 1}^n F_i$, since each $F_i$ must equal some $G_j$, so (A6)
proves (A5). 

In the two-sided case

\be
\prod_{i = 1}^n \frac{f_i}{F_i} = [\prod_{i = 1}^n
\frac{f_i}{G_{u_i}}] \ [\prod_{i = 1}^n \frac{G_{u_i}}{F_i}].
\ee

\noindent But the first bracketed factor on the right is the same as
$\prod_{i = 1}^n(f_i / F_i)$ for the one-sided case, and formula (4.7)
says that 

\be
\frac{G_{u_i}}{F_i} = \frac{G_{u_i}}{G_{u_i}-G_{v_i+}} = [1 -
\prod_{u_i}^{v_i} (1 - h_k)]^{-1} 
\ee

\noindent which together give the lemma (A4). Finally, differentiating
the log likelihood 

\be
\log \ L = \sum_{j = 1}^{n - 1} \ [\ \log \ h_j + (N_j - 1) \log \ (1 -
h_j)] - \sum_{i = 1}^n \ \log \ [1 - \prod_{u_i}^{v_i} (1 - h_k)]. 
\ee

\noindent gives

\be
\frac{\partial \log L}{\partial h_j} = \frac{1}{h_j} - \frac{1}{1 -
  h_j} \ [(N_j - 1) + \sum_{i : j \in R_i} \ \frac{G_{v_{i+}}}{F_i} \ ]
  \ ,
\ee
  
\noindent which is equivalent to the hazard rate theorem (4.8), (4.9).

The likelihood equality (A.5) for the one-sided case can be obtained
directly from familiar survival analysis arguments involving
successive conditionally independent binomial likelihoods. See section
2 of Efron (1988). In the two-sided case successive conditioning gives
the messier expression

\be
L = [ \ \prod_{j = 1}^n \prod_{i \in {\cal N}_j}(1 - h_{ij}) \ ] \
[ \ \prod_{j = 1}^n \ \frac{h_{jj}}{1 - h_{jj}} \ ] \ , 
\ee

\noindent when $h_{ij}$ is the hazard function for $y_i$, 

\be
h_{ij} = f_j \ / \sum_{j \leq k \leq v_i}f_k \ , 
\ee

\noindent and ${\cal N}_j = \{ i: \ u_i \leq j \quad {\rm and} \quad i
\geq j \}$, the ``$j^{th}$ risk set''. The lemma says that this
reduces to (A4).

\bigskip

\noindent {\bf Special Exponential Families}\quad Classic exponential
families such as the normal, Poisson, and binomial have mathematical
properties that greatly simplify their use. Modern computational
equipment allows us to design special exponential families (SEF) for
particular applications without worrying about mathematical
tractability. The SEF appearing in Figure 3 has a density of the form

\be
f_{\eta} (y) = e^{\eta^\prime t(y) - \phi(\eta)} \quad {\rm
  for} \quad y \ e \ {\cal Y}
\ee

\noindent where $t(y) = (y, y^2, y^3), \quad \eta = (\eta_1, \eta_2,
  \eta_3)$, ${\cal R} = [ \ \min \ (y_i), \max (y_i) \ ]$ and $\phi(\eta)$ is chosen to
  make $\int_{{\cal Y}} f_{\eta}(y) dy = 1$ 

The cubic SEF shown in Figure 3 used the MLE $\widehat \eta$ of the
parameter vector $\eta$. In order to calculate $\widehat \eta$ we
define

\be
T_{\eta}(g) = \int_{\cal R} g(y) e^{\eta^\prime t(y)}dy, 
\ee

\noindent where $g(y)$ may be a vector or matrix function of $y$ in
which case the integrals in (A.13) are carried out
component-wise. Then the score function $\dot \ell_\eta ({\bf y}) =
\frac{\partial}{\partial \eta} \ \log f_\eta ({\bf y})$ for the
truncated data structure (2.2), (2.3) is 

\be
\dot \ell_\eta ({\bf y}) = \sum_{i = 1}^n [ \ t(y_i) - T_\eta (I_i
\cdot t) \ / \ T_\eta (I_i)], 
\ee

\noindent where $I_i(y)$ is the indicator function for  $R_i$, as in
Efron and Tibshirani (1996). The second derivative matrix is given by

\be
-\ddot{\ell}_\eta({\bf y}) = \sum_{i = 1}^n \{ T_\eta (I_i \cdot t^2) \ / \
T_\eta (I_i) - [ \ T_\eta(I_i \cdot t) \ / \ T_\eta (I_i)) \ ]^2 \}, 
\ee

\noindent with $t^2$ indicating the $n \times n$ outer product matrix
$(t_i \ t_j)$. 

The MLE $\widehat \eta$, satisfying $\dot{\ell}_{\widehat \eta} (y) =
0$, is found by Newton-Raphson iteration 

\be
\widehat \eta (k + 1) - \widehat \eta (k) = [ \
-\ddot{\ell}_{\widehat \eta (k)} ({\bf y}) \ ]^{-1} \
\dot{\ell}_{\widehat \eta(k)}({\bf y}) \ , 
\ee

\noindent These calculations go quickly because they involve only
one-dimensional integrals (A.13), though many of them. Successive
models $t(y) = y_j \ t(y) = (y, y^2),\ldots$ were tried on the data
for Figure 2. Standard hypothesis tests led to the choice of a cubic
model. These tests used the estimated covariance matrix $[ \
-\ddot{\ell}_{\widehat \eta} ({\bf y}) \ ]^{-1}$ from (A.15) to assess
  the significance of the $\widehat \eta$ coefficients, e.g. whether or
  not $\widehat \eta_3$ is significantly non-zero for the cubic
  model. 

\noindent {\it Note}\quad The nonparametric MLE is itself an SEF
estimate, where the components of $t(y)$ in (A.12) are delta functions
on the observed $y_i$ values, $t(y) = (\delta (y - y_1), \ \delta (y -
y_2),\ldots, \delta (y - y_n)).$ 

\bigskip

\noindent {\bf Bootstrap and Permutation Calculations}\quad Section 5
discusses bootstrap approximations to permutation tests for the null
hypothesis of independence $H_o$. The situation there, involving
doubly truncated data, looks complicated but in fact it is perfectly
analogous to much simpler and more familiar statistical problems.

Consider the problem of testing the equality of two binomial
probabilities. We observe independent Bernoulli variates 

\begin{eqnarray}
y_{11}, y_{12}, \ldots y_{1n_1} \ \sim \ Bi(1, \pi_1) \qquad \qquad
\quad \quad \\
y_{21}, y_{22}, \ldots y_{2n_2} \ \sim \ Bi(1, \pi) \qquad \qquad
\qquad \end{eqnarray}

\noindent and wish to test the null hypothesis $H_o$ that $\pi_1 =
\pi_2 = \pi$ (say). Letting $x_1 = \sum_{i = 1}^{n_1} \ y_{1i} \quad
{\rm and} \quad x_2 = \sum_{i = 1}^{n_2} \ y_{2i}$ we can arrange the
data in a $2 \times 2$ table, first row $(x_1, n_1 - y_2)$ and second
row $(x_2, n_2 - x_2)$, in which case $H_o$ is the usual hypothesis of
independence for the table. 

If $H_o$ is true than $x_+ = x_1 + x_2$ is a sufficient statistic and
$x_+ \ \sim \ Bi(n_+, \pi)$ with $n_+ = n_1 + n_2$. Moreover all
$n_+$! permutations of the combined data vector ${\bf y} = (y_{11},
y_{12},\ldots, y_{2n_2})$ are equally likely given $x_+$, allowing us
to construct permutation tests for $H_o$ without worrying about the
nuisance parameter $\pi$. The permutation test based on the statistic 

\be
S = p_1 - p_2 \qquad \qquad [ \ p_1 = x_1 / n_1 \quad p_2 = x_2 /
n_2] 
\ee

\noindent is Fisher's exact test for independence. $S$ has permutation
expectation zero and variance

\be
\sigma_{\rm perm}^2 = \frac{n_+}{n_+-1} \widehat \pi (1 - \widehat
\pi)(\frac{1}{n_1} + \frac{1}{n_2}), \qquad [ \ \widehat \pi = x_+ /
n_+] 
\ee

Suppose we did not know formula (A20) and wished to approximate
$\sigma^2_{\rm perm}$ by bootstrap simulation. The null hypothesis
bootstrap replaces (A18) with 

\be
y_{11}^\ast, y_{12}^\ast, \ldots, y_{2n_2}^\ast \ \buildrel \rm
{\rm ind} \over \sim \ Bi(1, \widehat \pi), 
\ee

\noindent giving

\be
S^\ast = p_1^\ast - p_2^\ast = \sum_{i = 1}^{n_1} y_{1i}^\ast / n_1
- \sum_{i = 1}^{n_2} y_{2i}^\ast / n_2 . 
\ee

\noindent $S^\ast$ has bootstrap expectation zero and variance

\be
\widehat \sigma_{\rm perm}^2 = \widehat \pi (1 - \widehat \pi) \
(\frac{1}{n_1} + \frac{1}{n_2}). 
\ee

\noindent The usual $\chi^2$ statistic for testing independence is
exactly $S^2 / \widehat \sigma_{\rm perm}^2$. 

The bootstrap sampling is unconditional in the sense that the samples
do not necessarily have $x_+^\ast = x_+$. Nevertheless we get an
approximation that is accurate to second order,

\be
\widehat \sigma_{\rm perm}^2 / \sigma_{\rm perm}^2 = 1 + O(1 /
n_+). 
\ee

\noindent It can be shown that this is no accident, and that
second-order accuracy follows for any situation of the following sort:
\quad the observed data can be written as $(A, B)$ where under $H_o$, $A$
has an exponential family of densities and the conditional density
$f(B | A)$ does not depend on the parameters of the exponential
family. 

In the binomial situation $A = x_+$ and $B$ is a permutation of a
vector of $x_+$ 1's and $n_+ - x_+$ 0's. Under $H_o$, $A \sim Bi(n_+,
\pi)$ and $B | A$ is uniform. It is easiest to see the connection with
our truncated data problem through the SEF formulation. Formula (A.13)
leads to an exponential family of densities for ${\bf y}$ under the
truncated data model (2.2), with sufficient statistic $A = \sum_{i =
  1}^n t (y_i)$. As the exponential family (A.13) grows larger, say by
including more polynomial terms in $t(y) = (y, y^2, y^3, y^4,
\ldots)$, $A$ itself gets bigger, but we can always take $A = {\bf y}_{(
\ )}$, the order statistic of ${\bf y}$. But $A = {\bf y}_{( \ )}$ is
equivalent to taking $A = {\cal Y}$ (since the regions $R_i$ are known
ancillaries), as we did for the permutation tests of section 2.

\clearpage

\centerline {\bf References}

\noindent Efron, B. and Tibshirani R. (1996). ``Using specially designed
exponential families for density estimation.'' {\it Annals Stat} {\bf
  24}, p. 2431-2461.

\medskip

\noindent Efron, B. (1996). ``Empirical Bayes methods for combining
  likelihoods'' (with discussion) {\it Journal of American Statistical
  Association} {\bf 91}, p. 538-565.

\medskip

\noindent Efron, B. and Petrosian V. (1994), ``Survival analysis of
  the gamma-ray burst data'' {\it Journal of American Statistical
  Association} {\bf 89}, p. 452-462.

\medskip

\noindent Efron, B. and Petrosian V. (1992) ``A simple test of
independence for truncated data with applications to redshift
surveys'' {\it The Astrophysical Journal} {\bf 399}, p. 345-352.

\medskip

\noindent Efron, B. (1988) ``Logistic regression survival analysis,
and the Kaplan-Meier curve'' {\it Journal of American Statistical
  Association} {\bf 83}, p. 414-425.

\medskip

\noindent Gelman, A. and Rubin D. (1992) ``Iterative simulation using
single and multiple sequences'' (with discussion) {\it Statistical
  Science} {\bf 7}, p. 457-511.

\medskip

\noindent Gilks, W., Clayton D. et al. (1993) ``Modeling complexity:
applications of Gibbs sampling in medicine'' (with discussion) {\it
  Journal of the Royal Statistical Society} B {\bf 55}, p. 39-102.

\medskip

\noindent Lynden-Bell, D. (1971) ``A method of allowing for known
observational selection in small samples applied to 3CR quasars'' {\it
  Mon. Nat. R. Ast. Soc.} {\bf 155}, p. 95-118.

\medskip

\noindent McLaren C., Wagstaff M, Brittegram G., Jacobs A. (1991)
``Detection of two-component mixtures of lognormal distributions in
grouped, doubly truncated data analysis of red blood cell volume
distributions'' {\it Biometrics} {\bf 47}, p. 607-622.

\medskip

\noindent Romano, J. (1988) ``A bootstrap revival of some
nonparametric distance tests'' {\it Journal of American Statistical
  Association} {\bf 83}, p. 698-708.

\medskip

\noindent Tsai, W. (1990) ``Testing the independence of truncation
time and failure time'' {\it Biometrika} {\bf 77}, p. 169-77.

\medskip

\noindent Turnbull, B. (1996) ``The empirical distribution function
with arbitrarily grouped censored, and truncated data'' {\it Journal
  of the Royal Statistical Society} B{\bf 38}, p. 290-295.

\medskip

\noindent Weinberg, S. (1972) {\it Gravitation and Cosmology}, Wiley, New
York.

\end{document}